\documentclass[11pt]{article}
\usepackage{fullpage}
\usepackage{amsmath}
\usepackage{amsfonts}
\usepackage[dvips]{epsfig}

\def\##1{\underline{#1}}
\def\=#1{\underline{\underline{#1}}}

\def\+#1{\underline{\bf #1}}
\def\*#1{\underline{\underline{\bf #1}}}

\def\.{\mbox{ \tiny{$^\bullet$} }}
\def\I{\*I}
\def\curl{\nabla\times}

\def\eps{\epsilon}
\def\epso{\epsilon_{\scriptscriptstyle 0}}
\def\muo{\mu_{\scriptscriptstyle 0}}
\def\ko{k_{\scriptscriptstyle 0}}
\def\lo{\lambda_{\scriptscriptstyle 0}}

\def\c#1{\cite{#1}}
\def\l#1{\label{#1}}
\def\r#1{(\ref{#1})}

\def\le{\left(}
\def\ri{\right)}
\def\les{\left[}
\def\ris{\right]}
\def\lec{\left\{}
\def\ric{\right\}}

\begin{document}

\begin{center}
\Large{\bf {\LARGE Depolarization regions of nonzero volume
 in  bianisotropic homogenized composites }}

\normalsize \vspace{6mm}

Jiajia Cui\footnote{email:  s0457353@sms.ed.ac.uk} and  Tom G.
Mackay\footnote{  email: T.Mackay@ed.ac.uk}

\vspace{4mm}

\noindent{ \emph{School of Mathematics,  University of Edinburgh,\\
 Edinburgh EH9 3JZ,
United Kingdom.} }

\vspace{12mm}

{\bf Abstract}\end{center}

In conventional approaches to the homogenization of
 random particulate
composites, the component phase particles are often treated
mathematically as vanishingly small, point--like entities. The
electromagnetic responses of these component phase particles are
provided by depolarization dyadics which  derive from the
singularity of the corresponding dyadic Green functions. Through
neglecting the spatial extent of the depolarization region,
important information may be lost, particularly relating to coherent
scattering losses. We present an extension to the
strong--property--fluctuation theory in which depolarization regions
of nonzero volume and ellipsoidal geometry are accommodated.
Therein, both the size and spatial distribution of the component
phase particles are taken into account. The analysis is developed
within the most general linear setting of bianisotropic homogenized
composite mediums (HCMs). Numerical studies of the constitutive
parameters are presented for representative examples of HCM; both
Lorentz--reciprocal and Lorentz--nonreciprocal HCMs are considered.
These studies reveal that estimates of the HCM constitutive
parameters in relation to volume fraction, particle eccentricity,
particle orientation and correlation length are all significantly
influenced by the size of the component phase particles.

\vspace{8mm}





\section{Introduction}

The homogenization of a composite medium comprising two (or more)
component phases provides the backdrop for this study. The composite
may be  regarded  as an effectively homogenous medium as long as
wavelengths are sufficiently long compared with the
 dimensions of the component phase particles. In
electromagnetics, the  estimation of
   the constitutive parameters  of homogenized
composite mediums (HCMs) is a matter of long--standing, and ongoing,
scientific and technological importance  \c{L96}. Indeed, recent
developments  pertaining to HCM--based metamaterials serve to
 highlight the need for accurate formalisms in order to estimate
the constitutive parameters of complex HCMs \c{Walser,M05}.

In conventional approaches to homogenization the microstructural
details of the component phases are often inadequately incorporated
\c{Michel00}. For example,  the standard versions of the
widely--applied Maxwell Garnett \c{LW93} and Bruggeman \c{Ward}
formalisms utilize simplistic descriptions of the spatial
distributions and sizes of the component phase particles. An
alternative approach to homogenization is provided by the
 strong--property--fluctuation theory (SPFT), in which a comprehensive description
of the distributional statistics of the component phases can be
accommodated. While the origins of the
 SPFT lie in wave propagation studies for continuous
random mediums \c{Frisch,Ryzhov}, the SPFT has lately
  gained prominence  in
the homogenization of particulate composites
\c{TK81,Genchev,Z94,ML95,MLW00}. By means of the SPFT,  the HCM
constitutive parameters are estimated  as successive refinements to
the constitutive parameters of a homogenous comparison medium.
Iterates are expressed in terms of correlation functions describing
the spatial distributions of the component phases. In principle,
correlation functions of arbitrarily high order may be incorporated;
in practice, the SPFT is usually implemented at the second order
level of approximation. In fact, convergence of the SPFT scheme at
the second order level of approximation has been established for a
wide range of linear HCMs \c{MLW01}. Within the second order SPFT, a
two--point correlation function and its associated correlation
length characterize the component phase distributions. At lowest
order (i.e., zeroth and first order), the SPFT estimate of HCM
constitutive parameters is the same as  that of the Bruggeman
homogenization formalism \c{TK81,MLW00}.

The size of the component phase particles may be explicitly
incorporated within  homogenization formalisms via depolarization
dyadics \c{Michel00}. These dyadics are central to homogenization
analyses as they characterize the electromagnetic field inside
component phase particles embedded within a homogenous background.
Often the component phase particles are treated as vanishingly
small, point--like entities. Thereby, the corresponding
depolarization dyadic is represented by the singularity of the
associated dyadic Green function \c{M97,MW97}. However, potentially
important information is lost  through neglecting the spatial extent
of the component phase particles~---~especially if coherent
scattering losses are under consideration \c{Doyle,Dungey}. To
address this issue, extended versions of both the Maxwell Garnett
formalism \c{SL93a,SL93b,Prinkey} and the Bruggeman formalism
\c{Prinkey, S96} have been developed in which a nonzero volume is
attributed to the component phase particles. However, these analyses
apply only to isotropic HCMs and adopt a simplistic description of
 the distributional statistics of the
component phases. Recently, an extended version of the SPFT was
established for anisotropic dielectric HCMs, in which the sizes of
the component particles and their spatial distributions were taken
into account \c{M04}.

In the following sections, the second order SPFT for the most
general class of linear HCMs, namely bianistropic HCMs
\c{ML_Prog_Opt}, is extended to accommodate component phase
particles of nonzero size. The depolarization dyadic appropriate to
an ellipsoidal particle of nonzero size, embedded in a bianisotropic
medium, is developed in \S\ref{depol_section}. The incorporation of
this depolarization dyadic within the SPFT is then outlined in
\S\ref{homog_section}. The influence of the size of the component
phase particles upon the estimates of the HCM constitutive
parameters is explored in numerical studies in \S\ref{num_studies},
for both Lorentz--reciprocal and Lorentz--nonreciprocal HCMs.
Lastly, a few concluding remarks are provided in
\S\ref{conc_remarks}.

The following notation is adopted: Vector quantities are underlined.
Double underlining  and normal (bold) face signifies a 3$\times$3
(6$\times$6) dyadic.
 The inverse, adjoint, transpose  and determinant  of
a dyadic $\=M$ are denoted by $\=M^{-1}$, $\mbox{adj} \, \les \,\=M
\, \ris$, $\=M^{T}$ and $\mbox{det}\,\les \, \=M \, \ris $. The
3$\times$3 (6$\times$6) identity dyadic
 is represented by  $\,\=I\,$ ($\,\*I\,$).
All field--related quantities are implicitly functions of the
angular frequency $\omega$.
 The permittivity and
permeability of free space are denoted as $\epso$ and $\muo$,
respectively; the free-space wavenumber is $\ko = \omega \sqrt{
\epso \muo }\,$;  and $\lo = 2 \pi / \ko$. The real and imaginary
parts of $z \in \mathbb{C}$ are represented by $\mbox{Re} \,z$ and
$\mbox{Im} \,z$, respectively. A compact representation of the
constitutive parameters for the homogeneous bianisotropic medium
specified by the Tellegen constitutive relations \c{ML_Prog_Opt}
\begin{equation}
\left.
\begin{array}{l}
\#D(\#r) = \=\eps \. \#E(\#r) + \=\xi \. \#H(\#r) \\
\#B(\#r) = \=\zeta \. \#E(\#r) + \=\mu \. \#H(\#r)
\end{array}
\right\}
\end{equation}
is provided by  the 6$\times$6 constitutive dyadic
\begin{equation}
\*K_{\,} = \les \begin{array}{cc} \=\eps_{\, } & \=\xi_{\, }
\\ \=\zeta_{\,} & \=\mu_{\,} \end{array} \ris.
\end{equation}
Herein, $\=\eps_{\, }$ and $\=\mu_{\, }$ are the  3$\times$3
permittivity and  permeability dyadics , respectively, while
$\=\xi_{\, }$ and $\=\zeta_{\, }$ are the 3$\times$3
magneto\-electric dyadics. In the following, subscripts on $\*K$ are
used to identify the particular medium that $\*K$ describes.

\section{Depolarization region} \l{depol_section}

Let us consider an ellipsoidal particle of volume $V^{\eta}_e$,
oriented arbitrarily in $\mathbb{R}^3$. The ellipsoidal surface of $
V^{\eta}_e$ is parameterized
 by
\begin{equation}
\#r_{\,e}(\theta,\phi) = \eta \, \=U\.\#{\hat r} \, (\theta,
\phi),
\end{equation}
where  $\#{\hat r} \, (\theta, \phi)$ is the radial unit vector
specified by the spherical polar coordinates $\theta$ and $\phi$.
The 3$\times$3 shape dyadic $\=U$, which is real symmetric  with
unit determinant,
 maps the  spherical
region
 $V^\eta$ of radius $\eta$ onto the ellipsoidal region  $V^{\eta}_e$.
 The linear  dimensions of the ellipsoidal particle, as determined by $\eta$, are
assumed to be sufficiently small that the electromagnetic
long--wavelength regime pertains, but not vanishingly small.

Suppose now that the ellipsoidal particle is embedded within a
 bianisotropic comparison medium,
 characterized by  the 6$\times$6 constitutive dyadic $\*K_{\,cm} $.
 The comparison medium is homogeneous. The electromagnetic
response of the ellipsoidal particle is  provided by the
depolarization dyadic \c{MW97}
\begin{equation}
\*D (\eta) = \int_{V^{\eta}_e} \, \*G_{\,cm} (\#r) \; d^3 \#r \,=
\int_{V^\eta} \, \*G_{\,cm} (\=U \. \#r) \; d^3 \#r \,.
\l{depol_def}
\end{equation}
Herein,  $\*G_{\,cm} (\#r)$ is the 6$\times$6  dyadic Green function
of the comparison medium which satisfies the nonhomogenous vector
Helmholtz equation \c{MW97}
\begin{equation}
\les \*L(\nabla) +i \omega \*K_{\, cm}\ris \. \*G_{\,cm}(\#r -
\#r')
 = \I \, \delta(\#r - \#r'), \l{Helm}
\end{equation}
with the linear differential operator
\begin{equation}
\*L(\nabla) = \les \begin{array}{cc} \=0 & \curl\=I \\ -\curl\=I &
\=0 \end{array} \ris
 \end{equation}
  and $\delta(\#r - \#r')$ being
the Dirac delta function.

Explicit representations of  Green functions are  not generally
available for anisotropic and bianisotropic mediums \c{W93}.
However, it suffices for our present purposes to consider the
Fourier transform of $\*G_{\,cm} (\#r)$, namely
\begin{equation}
\*{\tilde{G}}_{\,cm} (\#q) = \int_{\#r} \*G_{\,cm} (\#r) \, \exp (-
i \#q \. \#r ) \; d^3 \#r \,,
\end{equation}
which is delivered from equation \r{Helm} as
\begin{equation}
\underline{\underline{\tilde{\bf G}}}_{\, cm}(\#q) = \frac{1}{i
\omega} \, \les \underline{\underline{\tilde{\bf A}}}_{\, cm}(\#q)
\ris^{-1},
\end{equation}
where
\begin{equation}
\underline{\underline{\tilde{\bf A}}}_{\, cm}(\#q)= \les
\begin{array}{cc} \=0 &  (\#q /\omega) \times \I \\ \vspace{-8pt} & \\
-(\#q /\omega) \times \I & \=0
\end{array} \ris +
\underline{\underline{\bf K}}_{\, cm}\,. \l{Gq_Ad}
\end{equation}
Thereby, equation \r{depol_def} yields
 \c{M97, MW97}
\begin{eqnarray}
\*D (\eta) &=&
 \frac{\eta}{2 \pi^2} \, \int_{\#q} \frac{1}{q^2} \, \le \frac{\sin
(q \eta )}{q \eta} - \cos ( q \eta) \ri \, \*{\tilde{G}}_{\,cm}
(\=U^{-1}\.\#q) \; d^3 \#q \, .
\end{eqnarray}

In order to  consider the depolarization dydic of an particle of
nonzero volume, we  express  $\=D (\eta) $ as the sum
\begin{equation}
\*D (\eta) = \*D^{>0} (\eta)  + \*D^{ 0} , \l{D_eta_def}
\end{equation}
The two terms  on the left side of \r{D_eta_def} are given by
\begin{eqnarray}
&& \*D^{>0} (\eta) =  \frac{\eta}{2 \pi^2} \, \int_{\#q}
\frac{1}{q^2} \, \le \frac{\sin (q \eta )}{q \eta} - \cos ( q
\eta) \ri \, \*{\tilde{G}}^{\eta}_{\,cm} (\=U^{-1}\.\#q)
\; d^3 \#q \, , \l{D0} \\
&& \*D^{ 0} =  \frac{\eta}{2 \pi^2} \, \int_{\#q} \frac{1}{q^2} \,
\le \frac{\sin (q \eta )}{q \eta} - \cos ( q \eta) \ri \,
\*{\tilde{G}}^{ \infty}_{\,cm} (\=U^{-1}\.\hat{\#q}) \; d^3 \#q \,
, \l{Dinf}
\end{eqnarray}
with
\begin{eqnarray}
 \underline{\underline{\tilde{\bf G}}}^{\eta}_{\,cm}(\=U^{-1}\.\#q)
&=& \underline{\underline{\tilde{\bf G}}}_{\,cm} (\=U^{-1}\.\#q) -
\underline{\underline{\tilde{\bf G}}}^{\infty}_{\,cm}
(\=U^{-1}\.\hat{\#q})\,, \l{G_Go_Gi}
\\
\underline{\underline{\tilde{\bf
G}}}^{\infty}_{\,cm}(\=U^{-1}\.\hat{\#q}) &=& \lim_{q\rightarrow
\infty} \;  \underline{\underline{\tilde{\bf G}}}_{\,cm}
(\=U^{-1}\.\#q) \\
&=&  \frac{1}{i \omega \, b( \theta, \phi )} \les
\begin{array}{ccc} \alpha_\mu ( \theta, \phi )\, \hat{\#q}\,\hat{\#q} && - \alpha_\zeta ( \theta, \phi ) \,
\hat{\#q}\,\hat{\#q}
\\ \vspace{-8pt} & \\
- \alpha_\xi ( \theta, \phi ) \, \hat{\#q}\,\hat{\#q} && \alpha_\eps
( \theta, \phi ) \, \hat{\#q}\,\hat{\#q}
\end{array} \ris, \nonumber \\ &&
\end{eqnarray}
wherein the scalars
\begin{equation}
\alpha_p ( \theta, \phi ) = \hat{\#q}\. \=U^{-1} \. \={p}_{\,cm}  \.
\=U^{-1}\.\hat{\#q}\,, \qquad \qquad (p = \eps, \zeta, \xi, \mu)
\end{equation}
and
\begin{equation}
b( \theta, \phi ) = \les \alpha_\eps ( \theta, \phi )\, \alpha_\mu (
\theta, \phi ) \ris - \les \alpha_\xi ( \theta, \phi )\,
\alpha_\zeta ( \theta, \phi )\ris .
\end{equation}
The volume integral \r{Dinf} simplifies to the $\eta$--independent
surface integral
  \c{M97, MW97}
\begin{equation} \l{dd_def}
\*D^0 = \frac{1}{4 \pi}\, \=U^{-1}\. \le \int^{2 \pi}_{\phi = 0}
\, \int^{\pi}_{\theta = 0} \,\underline{\underline{\tilde{\bf
G}}}^{\infty}_{\,cm}(\=U^{-1}\.\hat{\#q})\;\; \sin \theta \;
d\theta \; d\phi \ri \. \=U^{-1}.
\end{equation}
For certain Lorentz--reciprocal comparison mediums,  the volume
integral \r{D0} which yields $\*D^{>0}(\eta)$ may be reduced to  a
surface integral, but for a general bianisotropic comparison medium
no such simplifications are available. The integrals \r{D0} and
\r{dd_def}  may be evaluated using standard numerical techniques
\c{Fortran}.

The dyadic  $ \*D^{0}$ represents the depolarization contribution
arising from the vanishingly small region of volume $
\displaystyle{\lim_{\eta \rightarrow 0}} V^{\eta}_e $, whereas the
dyadic $ \*D^{>0}(\eta)$ provides the depolarization contribution
arising from the region of nonzero volume $ \le V^{\eta}_e -
 \displaystyle{\lim_{\eta \rightarrow 0}} V^{\eta}_e \ri
 $.
In homogenization studies, it is common practice to
neglect
 $ \*D^{>0} (\eta)$ and
assume that the depolarization dyadic is given by  $ \*D^{0}$ alone
\c{Michel00}. However, studies of  isotropic \c{Doyle, Dungey,
 SL93a,Prinkey, S96} and anisotropic \c{M04} HCMs  have emphasized the
importance of
 the nonzero spatial extent of depolarization
regions.

\section{Homogenization } \l{homog_section}

The SPFT may be implemented to estimate the constitutive parameters
of HCMs \c{TK81}. Let us consider the homogenization of a two--phase
composite
 wherein the two component phases, labelled as $a$ and $b$,
comprise  ellipsoidal particles of shape specified by $\=U$ and
linear dimensions specified by $\eta > 0$. A random distribution of
identically oriented particles is envisaged, such that all space $V$
is partitioned into parts $V_a$ and $V_b$ containing the phases
labelled $a$ and $b$, respectively. We consider the most general
linear scenario wherein the component phases $a$ and $b$ are taken
to be bianisotropic mediums with 6$\times$6 constitutive dyadics
$\*K_{\,a}$ and $\*K_{\,b}$, respectively.

The distributional statistics of the component phases are described
in terms of moments of the characteristic functions \c{MLW00}
\begin{equation}
\Phi_{ \ell}(\#r) = \left\{ \begin{array}{ll} 1, & \qquad \#r \in
V_{\, \ell},\\ & \qquad \qquad \qquad \qquad \qquad \qquad (\ell=a,b) . \\
 0, & \qquad \#r \not\in V_{\, \ell}, \end{array} \right.
\end{equation}
 The volume fraction of phase $\ell$, namely $f_\ell$ , is given by
the first statistical moment of
 $\Phi_{\ell}$ ;
 i.e., $\langle \, \Phi_{\ell}(\#r) \, \rangle = f_\ell$ .
 Clearly,
 $f_a + f_b = 1$.
The second statistical moment of $\Phi_{\ell}$
 provides a two--point covariance function.
We adopt the physically--motivated form \c{TKN82}
\begin{equation}
\langle \, \Phi_\ell (\#r) \, \Phi_\ell (\#r')\,\rangle =
\left\{
\begin{array}{lll}
\langle \, \Phi_\ell (\#r) \, \rangle \langle \Phi_\ell
(\#r')\,\rangle\,, & & \hspace{10mm}  | \, \=U^{-1}\. \le   \#r - \#r' \ri | > L \,,\\ && \hspace{25mm} \\
\langle \, \Phi_\ell (\#r) \, \rangle \,, && \hspace{10mm}
 | \, \=U^{-1} \. \le  \#r -
\#r' \ri | \leq L\,,
\end{array}
\right.
 \l{cov}
\end{equation}
where $L>0$ is the correlation length, which is taken to be much
smaller than the electromagnetic wavelengths. Over a range of
physically--plausible covariance functions, it has been shown that
the specific form of the covariance function has only a secondary
influence on SPFT estimates of HCM constitutive parameters
\c{MLW01b}.

\subsection{Zeroth order SPFT}

The $n$th  order SPFT estimate of the HCM constitutive dyadic,
namely $ \*K^{[n]}_{\,HCM}$,
 is based upon the
iterative refinement of the comparison medium constitutive dyadic,
namely $\*K_{\,cm}$. To zeroth order and first order, the SPFT
permittivity estimate is identical to the comparison medium
permittivity \c{MLW00}; i.e.,
\begin{equation}
\*K^{[0]}_{\,HCM} = \*K^{[1]}_{\,HCM}
 = \*K_{\,cm}.
\end{equation}
The well--known Bruggeman homogenization formalism provides the
estimate of $\*K_{\,cm}$ \c{MLW00}.
 That is, $\*K_{\,cm}$ emerges through
solving the nonlinear equations
\begin{equation}
f_a\,\mbox{\boldmath$\=\chi$}^{a/cm}  +
f_b\,\mbox{\boldmath$\=\chi$}^{b/cm}  = \*0\,, \l{Br}
\end{equation}
 wherein the polarizability density dyadics
\begin{eqnarray}
 \mbox{\boldmath$\=\chi$}^{\ell/cm} & =& - i \omega
\le\,\*K_{\,\ell}  - \*K_{\,cm}  \,\ri\.\les \,\*I + i \omega \*D
(\eta) \. \le\, \*K_{\,\ell}  - \*K_{\,cm}  \,\ri \ris^{-1},
 \qquad  (\ell = a,b).
\end{eqnarray}

\subsection{Second order SPFT}

The SPFT is most widely implemented at the second order level (also
known as the bilocal approximation) which provides the following
estimate of the HCM constitutive dyadic \c{MLW00}
\begin{equation}
 \*K^{[2]}_{\,HCM}  =
 \*K_{\,cm}  - \frac{1}{i \omega} \les \,\I +
\mbox{\boldmath$\={\Sigma}$}^{[2]}  \. \*D (\eta) \,\ris^{-1} \.
\mbox{\boldmath$\={\Sigma}$}^{[2]} . \l{KDy0}
\end{equation}
Thus, the particle size  $\eta$ influences $\*K^{[2]}_{\,HCM}$
directly through the depolarization dyadic $\*D (\eta) $ and
indirectly through
 the \emph{mass operator} \c{Frisch} dyadic term
\begin{equation}
\underline{\underline{\mbox{\boldmath$\Sigma$}}}^{[2]}  =  f_a f_b
\le \mbox{\boldmath$\=\chi$}^{a/cm} -
\mbox{\boldmath$\=\chi$}^{b/cm}
 \ri \.   \*D^{>0} (L)
 \. \le \mbox{\boldmath$\=\chi$}^{a/cm} -
\mbox{\boldmath$\=\chi$}^{b/cm}  \ri. \l{spv}
\end{equation}
Notice that the correlation length $L$~---~which plays a key role in
the second order SPFT~---~does not feature in the zeroth order SPFT.

\section{Numerical studies} \l{num_studies}

We now apply  the  theoretical results presented in Sections 2 and 3
to two specific bianisotropic homogenizations scenarios: in
\S\ref{biax_bian} a biaxial bianisotropic HCM is considered and in
\S\ref{FCM} a Faraday chiral medium \c{Engheta,WL98} is considered.
The HCM in \S\ref{biax_bian} is Lorentz--reciprocal \c{Krowne}
whereas the HCM in \S\ref{FCM} is not. Numerical studies are
presented for representative examples, in order to explore the
influence of $\eta$
 in relation to volume fraction, particle eccentricity, particle
orientation and correlation length. In view of the vast parameter
space associated with bianisotropic mediums, only an illustrative
selection of graphical results are provided here.

 The following calculations were carried using an angular
frequency $\omega = 2 \pi \times 10^{10}$ $\mbox{rad}\,
\mbox{s}^{-1}$. Hence, $\lo = 2 \pi / \ko =0.030 $ m.

\subsection{Biaxial bianisotropic HCM} \l{biax_bian}

The homogenization of (i) a biaxial dielectric medium described by
the constitutive dyadic
\begin{equation}
\*K_{\,a} = \les \begin{array}{cc} \epso \,\mbox{diag} \le
\eps^x_a, \eps^y_a, \eps^z_a \ri   & \=0
\\ \vspace{-8pt} & \\ \=0 & \muo \, \=I \end{array} \ris
\end{equation}
and (ii) an isotropic chiral medium described  by the constitutive
dyadic
\begin{equation}
\*K_{\,b} = \les \begin{array}{cc} \epso \eps_b \, \=I & i
\sqrt{\epso \muo} \, \xi_b \, \=I
\\ \vspace{-8pt} & \\ - i
\sqrt{\epso \muo} \,\xi_b \, \=I &  \muo \, \=I\end{array} \ris
\l{iso_chiral}
\end{equation}
is investigated.
 The constitutive parameters selected for calculations are:
$\eps^x_a = 4 + i 0.12$, $\eps^y_a = 3 + i 0.1$, $\eps^z_a = 1.5 + i
0.08$; $\eps_b = 2.5 + i 0.1$, $\xi_b = 1 + i 0.07$ and $\mu_b =
1.75 + i 0.09$.

The shape dyadic of the constituent particles is taken to be
\begin{equation}
\=U = \frac{1}{\sqrt[3]{U_x U_y U_z}} \; \=R_{\,z} (\varphi) \.
\les \, \mbox{diag} (U_x, U_y, U_z) \ris \. \=R^T_{\,z} (\varphi),
 \l{U}
\end{equation}
with
\begin{equation}
\=R_{\,z} (\varphi) =
\le \begin{array}{ccc} \cos \varphi &  \sin \varphi & 0 \\
- \sin \varphi & \cos \varphi & 0 \\
0 & 0 & 1 \end{array} \ri.
\end{equation}
Thus, the principal axes of the ellipsoidal particles lie in the
$xy$ plane rotated by an angle $\varphi$, and  along the $z$ axis.
The shape parameters selected for calculations are: $U_x = 1 +
\rho$, $U_y = 1$ and $U_z = 1 - 0.5 \rho $; the eccentricity of the
ellipsoids is varied through the parameter $\rho$.

The corresponding HCM is a Lorentz--reciprocal, biaxial,
bianisotropic medium. In this case, the  volume integral
 in \r{D0} for $\*D^{>0}(\eta)$
is analogous to one which arises in the development of the second
order SPFT \c{MLW00}.  Thus, we may express $\*D^{>0}(\eta)$ as the
 surface integral
\begin{eqnarray}
\*D^{>0}(\eta) &=& \frac{\omega^3}{8 \pi i} \int^{2 \pi}_{\phi = 0}
\int^{\pi}_{\theta = 0} \Bigg( \frac{1}{\kappa_+ - \kappa_-} \Big\{
\frac{e^{i \eta q}}{q^2} \le 1 - i \eta  q\ri \Big[
\,\*N(\=U^{-1}\.\#q) \nonumber \\ && + \*N(-\=U^{-1}\.\#q) \Big]
\Big\}^{q= \sqrt{\kappa_+}}_{q=\sqrt{\kappa_-}} + \frac{2}{\kappa_+
\kappa_-}\, \*N(\#0) \Bigg)\, \sin \theta \; d \theta \; d \phi\,,
\end{eqnarray}
with
\begin{equation}
\*N(\=U^{-1}\.\#q) = \frac{1}{b(\theta, \phi)}\, \lec  \mbox{adj}
\les \underline{\underline{\tilde{\bf A}}}_{\, cm}(\=U^{-1}\.\#q)
\ris - \mbox{det} \les \underline{\underline{\tilde{\bf A}}}_{\,
cm}(\=U^{-1}\.\#q) \ris \underline{\underline{\tilde{\bf
G}}}^\infty_{\, cm}(\=U^{-1}\.\hat{\#q}) \ric
\end{equation}
and $\kappa_\pm$ being the $q^2$ roots of $\mbox{det} \les
\underline{\underline{\tilde{\bf A}}}_{\, cm}(\=U^{-1}\.\#q) \ris$.

\subsubsection{Particle size and volume fraction}

To focus upon the effect of particle size  $\eta$ in relation to
volume fraction $f_a$, we set the eccentricity $\rho = 0$, the
orientation angle $\varphi = 0$ and the correlation length $L = 0$.
The corresponding HCM constitutive dyadic has the form
\begin{equation}
\*K_{\,HCM} = \les \begin{array}{cc} \epso \,\mbox{diag} \le
\eps^x_{HCM}, \eps^y_{HCM}, \eps^z_{HCM} \ri   & i \sqrt{\epso
\muo} \,\mbox{diag} \le \xi^x_{HCM}, \xi^y_{HCM}, \xi^z_{HCM} \ri
\\ \vspace{-8pt} & \\ - i \sqrt{\epso
\muo} \,\mbox{diag} \le \xi^x_{HCM}, \xi^y_{HCM}, \xi^z_{HCM} \ri
& \muo  \,\mbox{diag} \le \mu^x_{HCM}, \mu^y_{HCM}, \mu^z_{HCM}
\ri
\end{array} \ris. \l{K_HCM_biax1}
\end{equation}

\begin{figure}[!ht]
\centering \psfull \epsfig{file=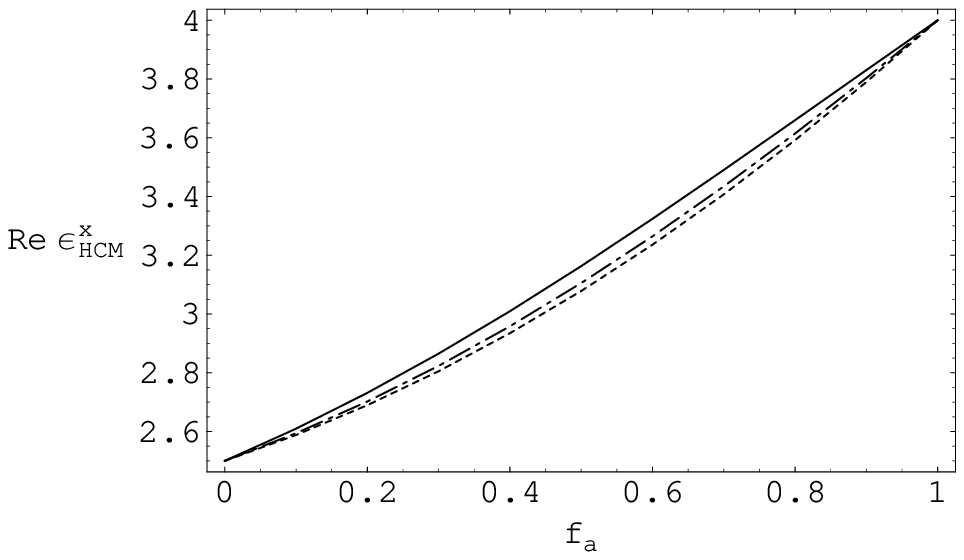,width=3.0in} \hfill
  \epsfig{file=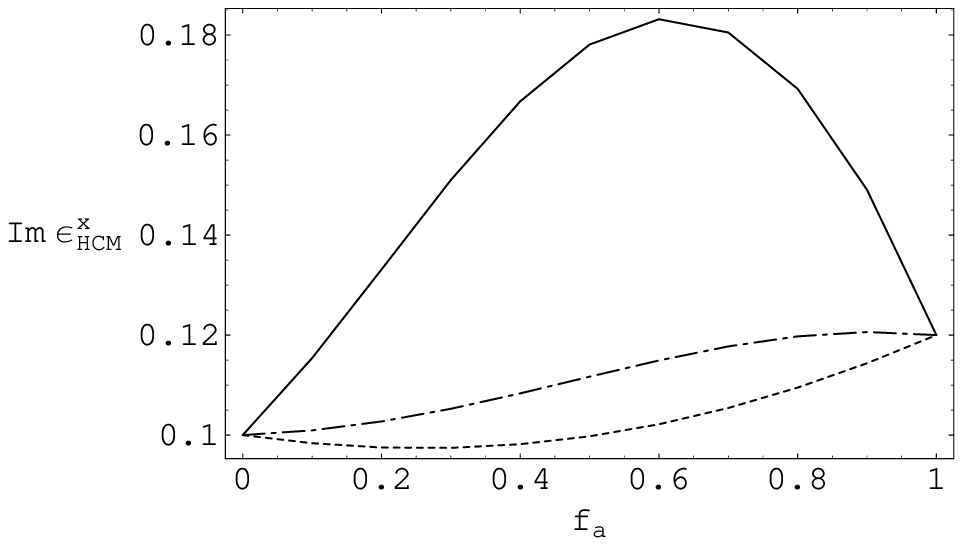,width=3.0in} \\
   \epsfig{file=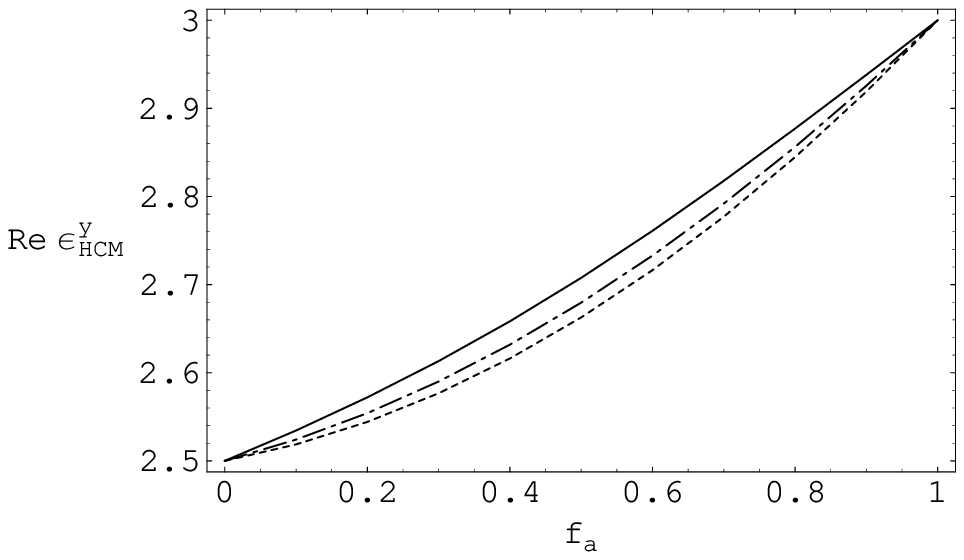,width=3.0in} \hfill
  \epsfig{file=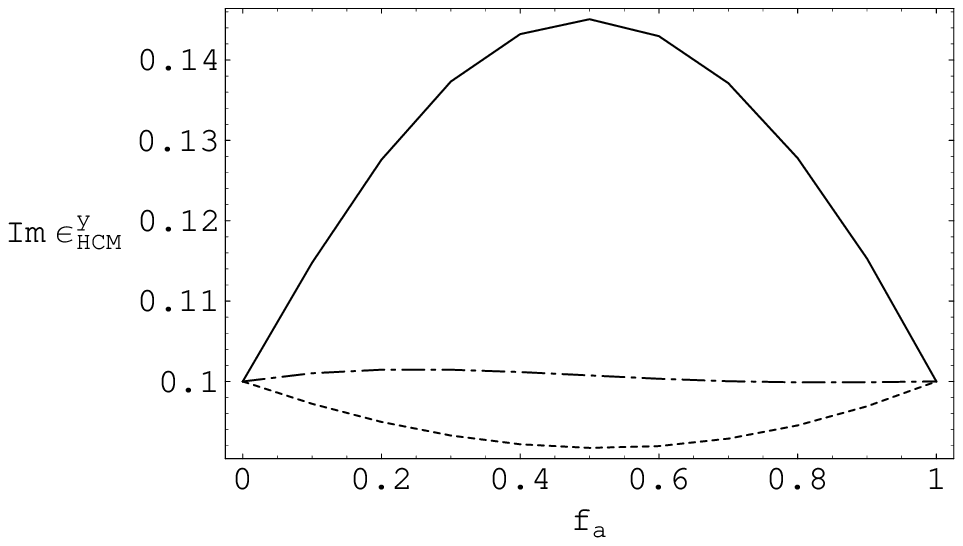,width=3.0in}\\
   \epsfig{file=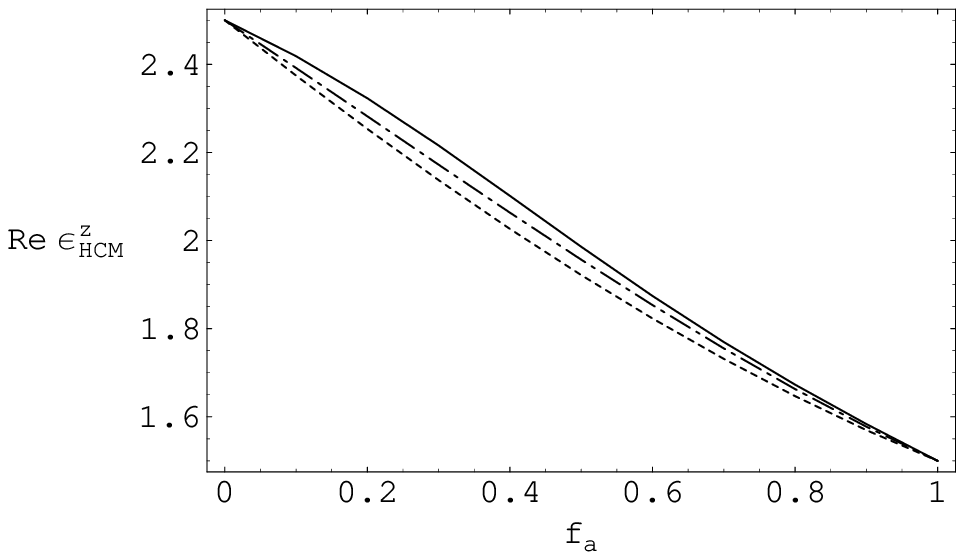,width=3.0in} \hfill
  \epsfig{file=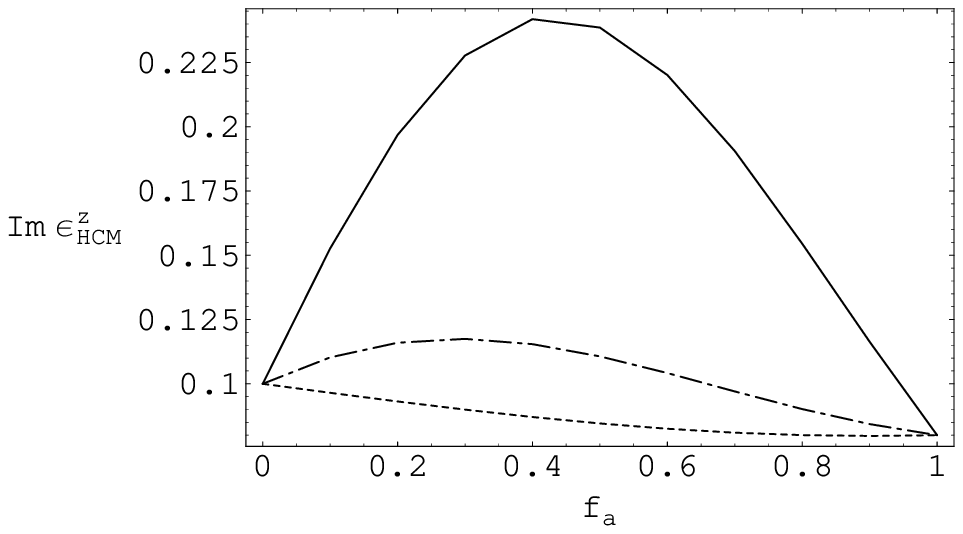,width=3.0in}
\caption{ \label{fig1} Real (left) and imaginary (right) parts of
the HCM constitutive parameters  $\eps^{x,y,z}_{HCM}$  plotted
against
 volume fraction $f_a \in (0,1) $ for $ \eta / \lo = 0$ (dashed curves), $ \eta / \lo =   0.05$ (broken dashed curves)
 and $ \eta / \lo =  0.1 $ (solid curves). The HCM is a biaxial bianisotropic medium. }
\end{figure}

In Figure~\ref{fig1}, the real and imaginary parts of
$\eps^{x,y,z}_{HCM}$ are plotted against
 $f_a \in (0,1) $ for $ \eta / \lo \in \lec 0, 0.05, 0.1 \ric$.
Notice that the HCM parameters are constrained to coincide with
those of component phase $b$ in the limit $f_a \rightarrow 0$, and
those of component phase $a$ in the limit $f_a \rightarrow 1$. The
influence of $\eta$ is more obviously observed on the imaginary
parts of $\eps^{x,y,z}_{HCM}$ than on the real parts. Indeed, the
imaginary parts of $\eps^{x,y,z}_{HCM}$ for $\eta / \lo = 0.1 $ at
mid--range values of $f_a$ are approximately twice as large as they
are for $\eta / \lo = 0$.
 The corresponding graphs
 for the HCM magnetoelectric parameters  $\xi^{x,y,z}_{HCM}$ and
 permeability parameters $\mu^{x,y,z}_{HCM}$ are qualitatively
 similar to those presented for the permeability parameters in Figure~\ref{fig1}.

\subsubsection{Particle size  and particle
eccentricity}

Next we turn to the effect of particle size  $\eta$ in relation to
particle eccentricity, as specified by $\rho$.  The volume fraction
is fixed at $f_a =  0.5$, the  orientation angle at $\varphi = 0$
and the correlation length at $L = 0$. The corresponding HCM
constitutive dyadic has the form \r{K_HCM_biax1}.

\begin{figure}[!ht]
\centering \psfull \epsfig{file=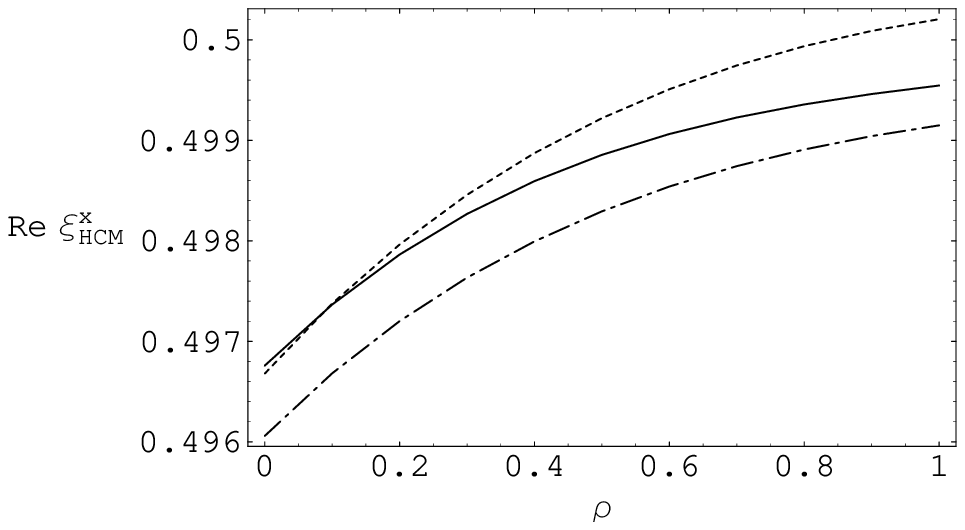,width=3.0in} \hfill
  \epsfig{file=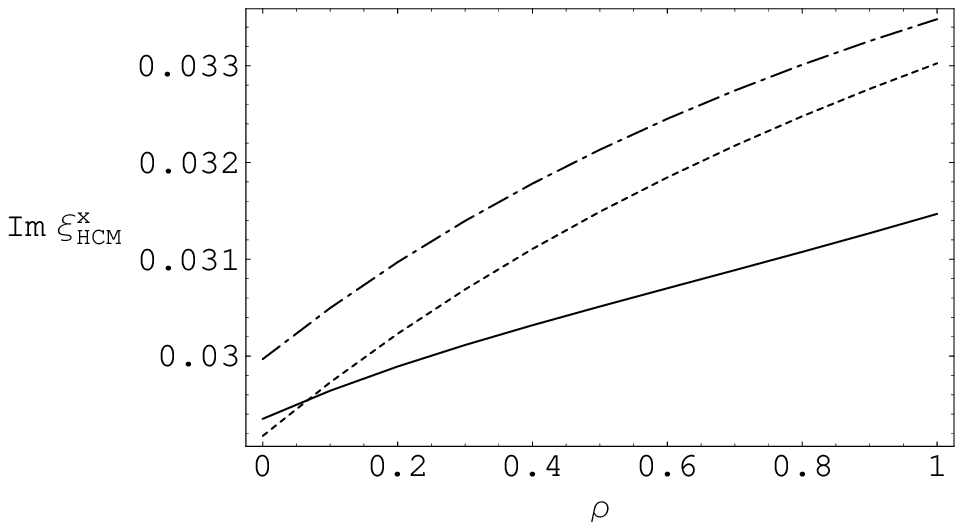,width=3.0in}
\caption{ \label{fig2} Real (left) and imaginary (right) parts of
the HCM constitutive parameter $\xi^{x}_{HCM}$  plotted against the
eccentricity parameter $\rho \in (0,1)$  for $ \eta / \lo = 0$
(dashed curves), $ \eta / \lo =   0.05$ (broken dashed curves)
 and $ \eta / \lo =  0.1 $ (solid curves). The HCM is a biaxial bianisotropic medium.
}
\end{figure}

 As a typical
example of the behaviour of HCM constitutive parameters,
 the real and imaginary parts of $\xi^{x}_{HCM}$
are graphed versus $\rho \in (0,1)$ for $ \eta / \lo \in \lec 0,
0.05, 0.1 \ric$ in Figure~\ref{fig2}. Regardless of the value of
$\eta$, the constitutive parameters vary
substantially~---~particularly their imaginary parts~---~as $\rho$
increases. Furthermore, there are significant differences  in  the
plots of $\xi^{x}_{HCM}$ presented for the three values of $\eta$.
The most striking differences are observed in the plots of the
imaginary parts of $\xi^{x}_{HCM}$. The corresponding graphs for the
constitutive parameters not presented in Figure~\ref{fig2} (i.e.,
 $\eps^{x,y,z}_{HCM}$, $\xi^{y,z}_{HCM}$ and
$\mu^{x,y,z}_{HCM}$) are broadly similar to those given in
Figure~\ref{fig2}.

\subsubsection{Particle size  and  particle
 orientation}

In order to investigate the effect of particle size  $\eta$ in
relation to particle orientation,  we fix the volume fraction $f_a =
0.5$, particle eccentricity $\rho = 1$ and the correlation length $L
= 0$. The resulting HCM has a constitutive dyadic of the form
\begin{equation}
\*K_{\,HCM} = \les \begin{array}{cc} \epso \, \le
\begin{array}{ccc} \eps^x_{HCM} & \eps^t_{HCM} & 0 \\  \vspace{-12pt}
&& \\
\eps^t_{HCM} & \eps^y_{HCM} & 0 \\ \vspace{-12pt}
&& \\
0 & 0 & \eps^z_{HCM}
\end{array}
\ri &
 i \sqrt{\epso \muo} \,
  \le
\begin{array}{ccc} \xi^x_{HCM} & \xi^t_{HCM} & 0 \\ \vspace{-12pt}
&& \\
\xi^t_{HCM} & \xi^y_{HCM} & 0 \\ \vspace{-12pt}
&& \\
0 & 0 & \xi^z_{HCM}
\end{array}
\ri
\\ \vspace{-8pt} & \\ - i \sqrt{\epso
\muo} \,\le
\begin{array}{ccc} \xi^x_{HCM} & \xi^t_{HCM} & 0 \\ \vspace{-12pt}
&& \\
\xi^t_{HCM} & \xi^y_{HCM} & 0 \\ \vspace{-12pt}
&& \\
0 & 0 & \xi^z_{HCM}
\end{array}
\ri & \muo \,\le
\begin{array}{ccc} \mu^x_{HCM} & \mu^t_{HCM} & 0 \\
\vspace{-12pt}
&& \\
\mu^t_{HCM} & \mu^y_{HCM} & 0 \\
\vspace{-12pt}
&& \\
0 & 0 & \mu^z_{HCM}
\end{array}
\ri
\end{array} \ris. \l{K_HCM_biax2}
\end{equation}

\begin{figure}[!ht]
\centering \psfull
   \epsfig{file=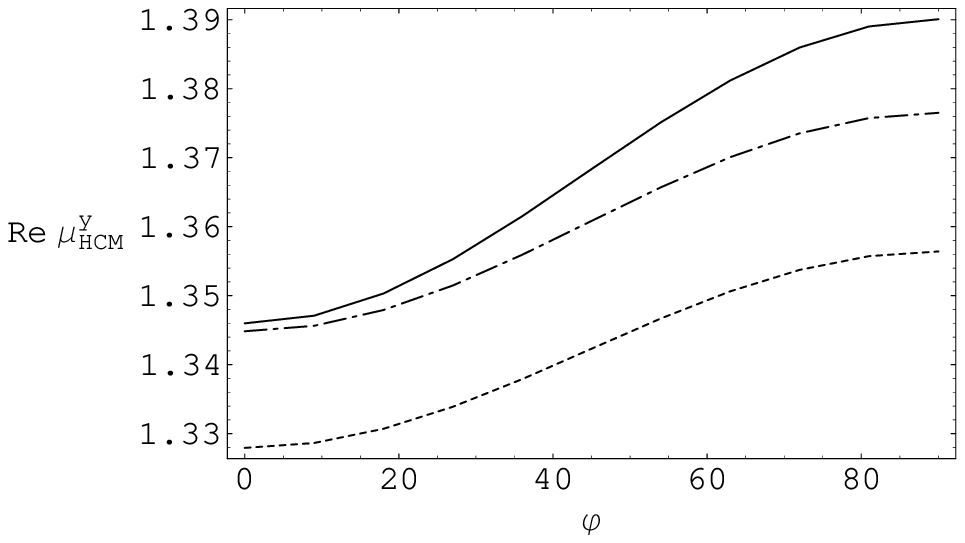,width=3.0in} \hfill
  \epsfig{file=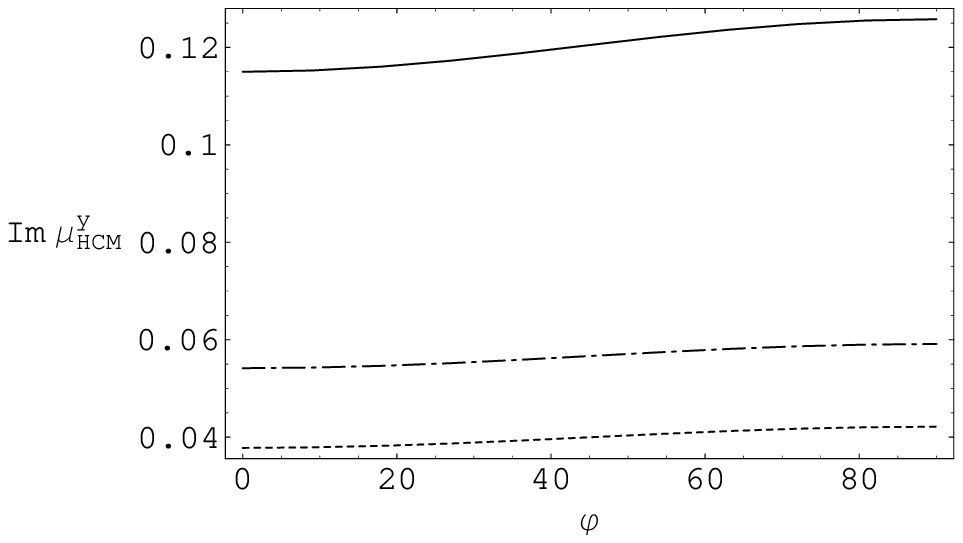,width=3.0in}\\
   \epsfig{file=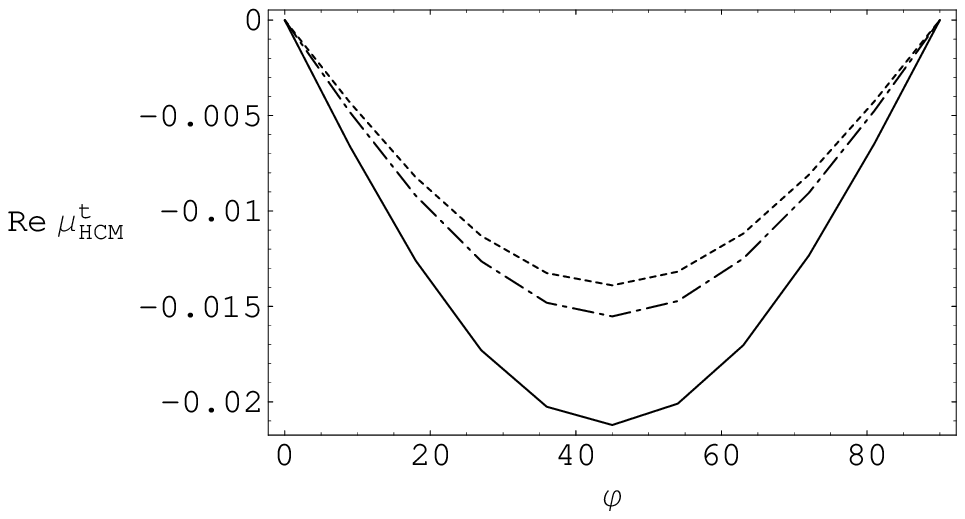,width=3.0in} \hfill
  \epsfig{file=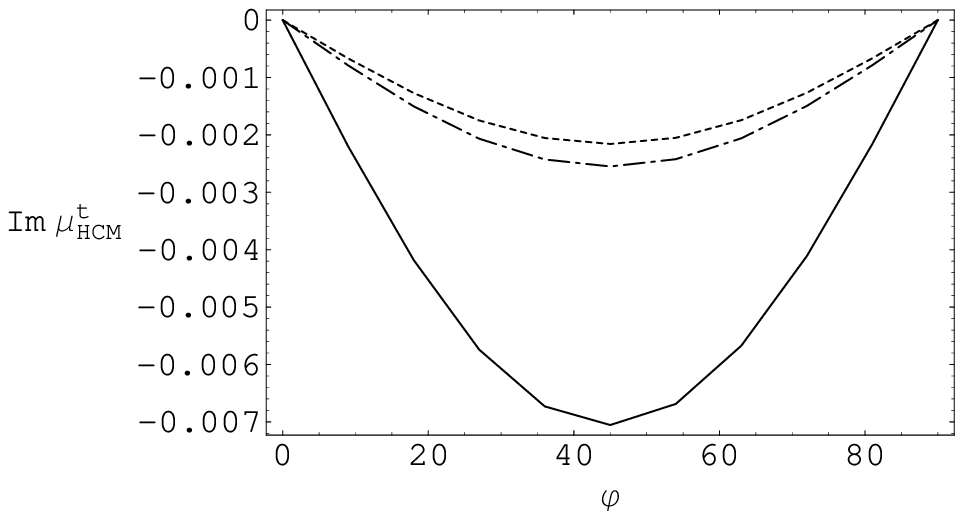,width=3.0in}
\caption{ \label{fig3} Real (left) and imaginary (right) parts of
the HCM constitutive parameters  $\mu^{y,t}_{HCM}$  plotted against
 orientation angle $\varphi \in (0,\pi/2)$ for $ \eta / \lo = 0$ (dashed curves), $ \eta / \lo =   0.05$ (broken dashed curves)
 and $ \eta / \lo =  0.1 $ (solid curves). The HCM is a biaxial bianisotropic medium.
}
\end{figure}

Illustrative numerical results are displayed in  Figure~\ref{fig3},
wherein the real and imaginary parts of $\mu^{y,t}_{HCM}$ are
plotted against $\varphi \in (0,\pi/2)$ for $ \eta / \lo \in \lec 0,
0.05, 0.1 \ric$. The off--diagonal constitutive parameter $\mu^t$
vanishes in the limits $\varphi \rightarrow 0$ and $\pi/2$ (as do
$\eps^t$ and $\xi^t$). Both the  real and imaginary parts of $\mu^t$
are  strongly influenced  by the particle size $\eta$, especially
for mid--range values of $\varphi$. The diagonal constitutive
parameter $\mu^y$ is also clearly sensitive to $\eta$. In the case
of $\mu^y$, the differences in behaviour for the three values of
$\eta$ are most apparent as $\varphi$ approaches 0 and $\pi/2$. The
graphs of $\mu^t$ are symmetric about $\varphi = \pi /4$, but those
of $\mu^y$ are not.
 The
HCM constitutive parameters that are not represented in
Figure~\ref{fig3} (i.e., $\eps^{x,y,z,t}_{HCM}$,
$\xi^{x,y,z,t}_{HCM}$ and $\mu^{x,z}_{HCM}$ ) exhibit behaviour with
respect to $\varphi$ which is generally similar to that exhibited by
$\mu^{y,t}_{HCM}$ in Figure~\ref{fig3}.

\subsubsection{Particle size and correlation length}

Lastly in this section,   particle size $\eta$ is considered in
relation to correlation length $L$. To do so, the following
parameters are fixed: volume fraction $f_a = 0.5$, orientation angle
$\varphi = 0$ and the eccentricity $\rho = 0$. The constitutive
dyadic of the HCM which arises has the form \r{K_HCM_biax1}.

\begin{figure}[!ht]
\centering \psfull \epsfig{file=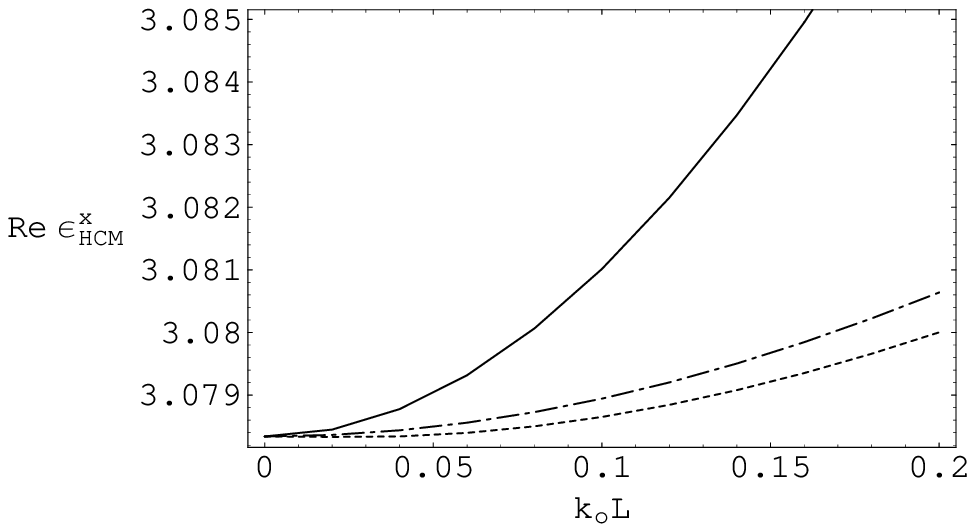,width=3.0in} \hfill
  \epsfig{file=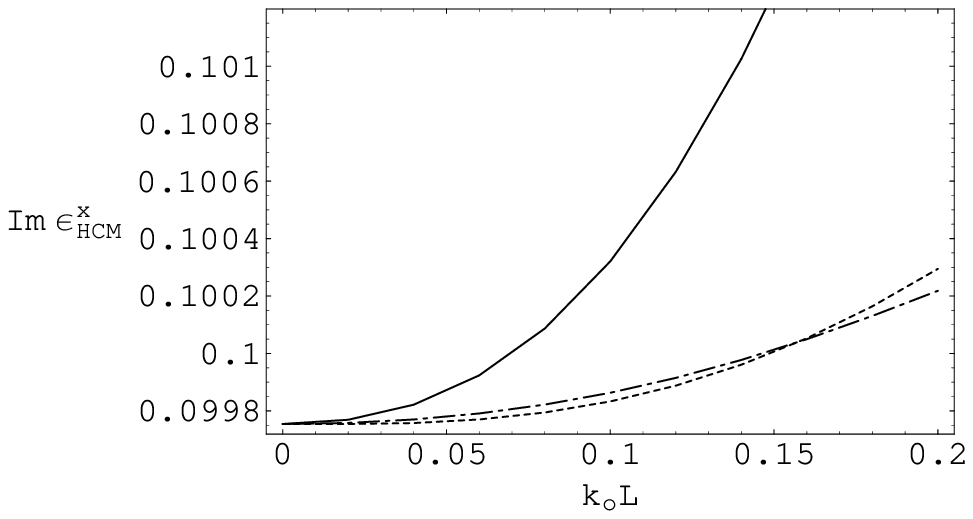,width=3.0in}
\caption{ \label{fig4} Real (left) and imaginary (right) parts of
the HCM constitutive parameter $\eps^{x}_{HCM}$  plotted against
 relative correlation length
  $\ko L  \in (0,0.2)$ for $ \eta / L = 0$ (dashed curves), $ \eta / L =   0.5$ (broken dashed curves)
 and $ \eta / L =  0.95 $ (solid curves). The HCM is a biaxial bianisotropic medium.
}
\end{figure}

In Figure~\ref{fig4}, graphs of the real and imaginary parts of
$\eps^{x}_{HCM}$ versus $\ko L  \in (0,0.2)$ are provided for $ \eta
/ L \in \lec 0, 0.5, 0.95 \ric$. It is clear that the imaginary part
of $\eps^{x}_{HCM}$ is strongly affected by increasing $L$; the real
part of $\eps^{x}_{HCM}$ is also affected but to a lesser degree.
Furthermore,  $\eps^{x}_{HCM}$ is much more sensitive  to $L$ at
larger values of $\eta$. The behaviour observed in Figure~\ref{fig4}
for $\eps^{x}_{HCM}$ with respect to $L$  is also generally observed
in the HCM constitutive parameters $\eps^{y,z}_{HCM}$,
$\xi^{x,y,z}_{HCM}$ and $\mu^{x,y,z}_{HCM}$ which are not
represented in Figure~\ref{fig4}.

\subsection{Faraday chiral medium} \l{FCM}

For our second homogenization scenario, we explore the
homogenization of (i) a gyrotropic magnetic medium described by the
constitutive dyadic
\begin{equation}
\*K_{\,a} = \les \begin{array}{cc} \epso \, \eps_a \, \=I   & \=0
\\ \vspace{-8pt} & \\ \=0 & \muo \,
\le
\begin{array}{ccc} \mu^x_a & i \mu^g_a & 0 \\
\vspace{-12pt}
&& \\
-i \mu^g_a & \mu^x_a & 0 \\
\vspace{-12pt}
&& \\
0 & 0 & \mu^z_a
\end{array}
\ri
 \end{array} \ris
\end{equation}
and (ii) an isotropic chiral medium described by the constitutive
dyadic \r{iso_chiral}. The constitutive parameters selected for
calculations are: $\eps_a = 1.2 + i 0.02$, $\mu^x_a = 3.5 + i 0.08$,
$\mu^g_a = 1.8 + i 0.05$, $\mu^z_b = 1.4 + i 0.04$; $\eps_b = 2.5 +
i 0.1$, $\xi_b = 1 + i 0.07$ and $\mu_b = 1.75 + i 0.09$. As in
\S\ref{biax_bian}, the shape dyadic of the constituent particles is
taken to have the form \r{U}, with the shape parameters selected for
calculations being: $U_x = 1 + \rho$, $U_y = 1$ and $U_z = 1 - 0.5
\rho $.

The HCM which results is a Faraday chiral medium
\c{Engheta,WL98,WLM98}. A HCM of the same form also arises from the
homogenization of a magnetically--biased plasma and an isotropic
chiral medium \c{MW01}.

\subsubsection{Particle size  and  volume fraction}

We begin by considering  the effect of particle size $\eta$ in
relation to volume fraction $f_a$. Accordingly,  the eccentricity is
fixed at $\rho = 0$, the orientation angle at $\varphi = 0$ and the
correlation length at $L = 0$. The HCM constitutive dyadic has the
form
\begin{equation}
\*K_{\,HCM} = \les \begin{array}{cc} \epso \,  \le
\begin{array}{ccc} \eps^x_{HCM} & i \eps^g_{HCM} & 0 \\
\vspace{-12pt}
&& \\
-i \eps^g_{HCM} & \eps^x_{HCM} & 0 \\
\vspace{-12pt}
&& \\
0 & 0 & \eps^z_{HCM}
\end{array}
\ri
 & i \sqrt{\epso
\muo} \,
  \le
\begin{array}{ccc} \xi^x_{HCM} & i \xi^g_{HCM} & 0 \\
\vspace{-12pt}
&& \\
-i \xi^g_{HCM} & \xi^x_{HCM} & 0 \\
\vspace{-12pt}
&& \\
0 & 0 & \xi^z_{HCM}
\end{array}
\ri
\\ \vspace{-8pt} & \\ - i \sqrt{\epso
\muo} \,
  \le
\begin{array}{ccc} \xi^x_{HCM} & i \xi^g_{HCM} & 0 \\
\vspace{-12pt}
&& \\
-i \xi^g_{HCM} & \xi^x_{HCM} & 0 \\
\vspace{-12pt}
&& \\
0 & 0 & \xi^z_{HCM}
\end{array}
\ri & \muo \,
  \le
\begin{array}{ccc} \mu^x_{HCM} & i \mu^g_{HCM} & 0 \\
\vspace{-12pt}
&& \\
-i \mu^g_{HCM} & \mu^x_{HCM} & 0 \\
\vspace{-12pt}
&& \\
0 & 0 & \mu^z_{HCM}
\end{array}
\ri
\end{array} \ris. \l{K_HCM_fcm1}
\end{equation}

\begin{figure}[!ht]
\centering \psfull \epsfig{file=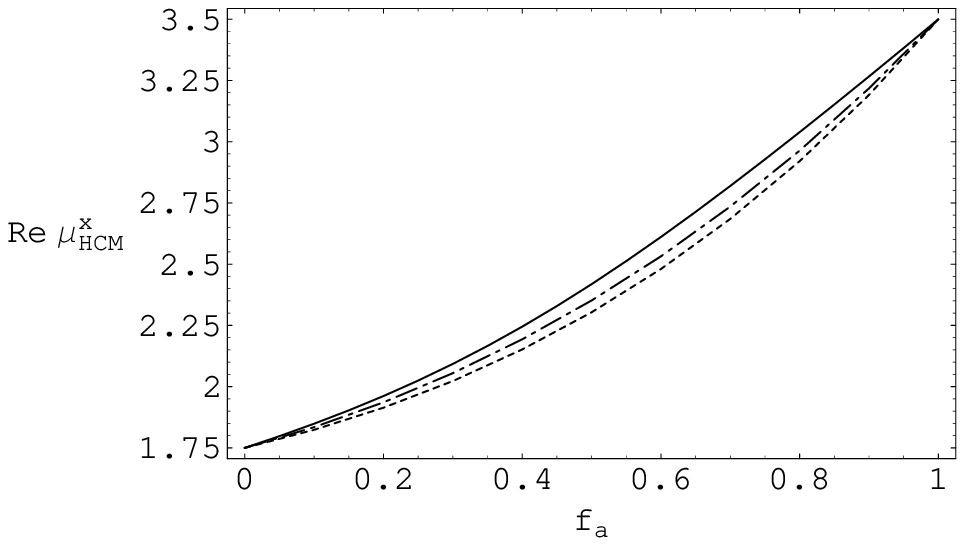,width=3.0in} \hfill
  \epsfig{file=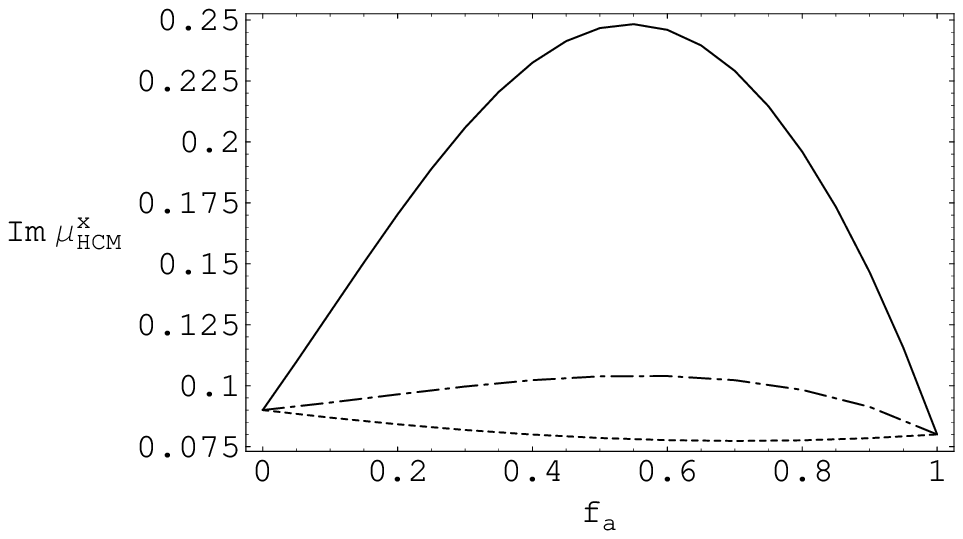,width=3.0in} \\
   \epsfig{file=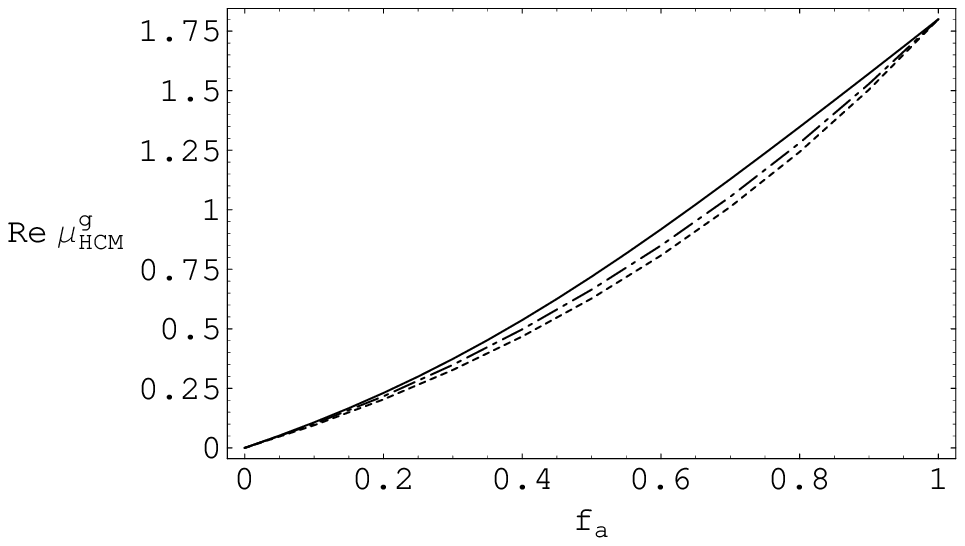,width=3.0in} \hfill
  \epsfig{file=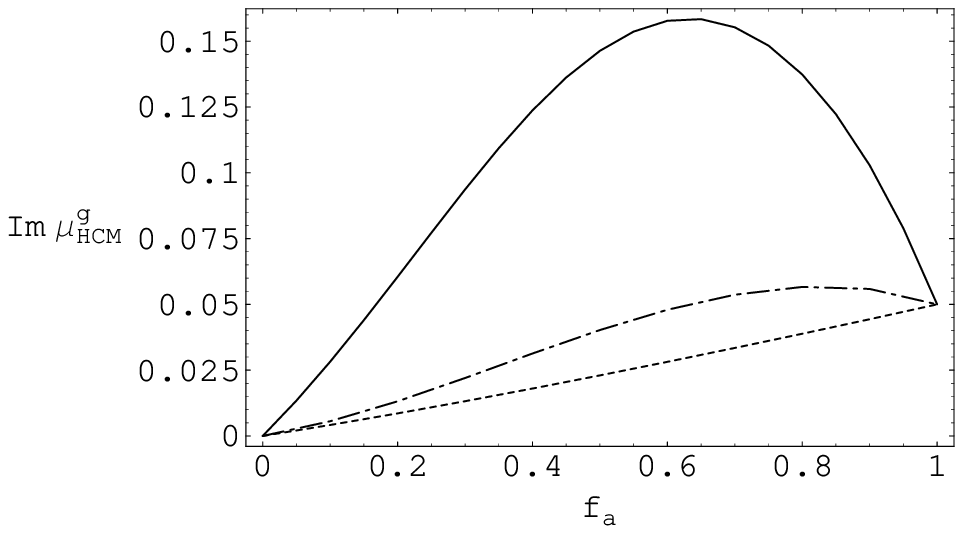,width=3.0in}
\caption{ \label{fig5} Real (left) and imaginary (right) parts of
the HCM constitutive parameters  $\mu^{x,g}_{HCM}$  plotted against
 volume fraction $f_a \in (0,1) $ for $ \eta / \lo = 0$ (dashed curves), $ \eta / \lo =   0.05$ (broken dashed curves)
 and $ \eta / \lo =  0.1 $ (solid curves). The HCM is a Faraday
 chiral medium.
}
\end{figure}

In Figure~\ref{fig5}, the real and imaginary parts of
$\mu^{x,g}_{HCM}$ are plotted against
 $f_a \in (0,1) $ for $ \eta / \lo \in \lec 0, 0.05, 0.1 \ric$.
As is the case in Figure~\ref{fig1}, the HCM constitutive parameters
are constrained such that they coincide with those of component
phase $b$ and $a$ in the limits $f_a \rightarrow 0$ and $1$,
respectively. The effect of $\eta$ on the real parts of
$\mu^{x,g}_{HCM}$ are relatively modest. In contrast,  $\eta$ has a
profound effect on the imaginary parts of $\mu^{x,g}_{HCM}$,
especially for mid--range values of $f_a$. The pattern of behaviour
presented in Figure~\ref{fig5} for the HCM permeability parameters
$\mu^{x,g}_{HCM}$ is mirrored by the HCM permittivity parameters
$\eps^{x,z,g}_{HCM}$ and magnetoelectric parameters
$\xi^{x,z,g}_{HCM}$, as well as $\mu^{z}_{HCM}$,  which are not
displayed in Figure~\ref{fig5}.

\subsubsection{Particle size and particle
eccentricity}

The effect of particle size  $\eta$ in relation to particle
eccentricity is considered next.  We set the volume fraction $f_a =
0.5$,  orientation angle $\varphi = 0$ and the correlation length $L
= 0$.  The HCM constitutive dyadic then has the form
\begin{equation}
\*K_{\,HCM} = \les \begin{array}{cc} \epso \,  \le
\begin{array}{ccc} \eps^x_{HCM} & i \eps^g_{HCM} & 0 \\
\vspace{-12pt}
&& \\
-i \eps^g_{HCM} & \eps^y_{HCM} & 0 \\
\vspace{-12pt}
&& \\
0 & 0 & \eps^z_{HCM}
\end{array}
\ri
 & i \sqrt{\epso
\muo} \,
  \le
\begin{array}{ccc} \xi^x_{HCM} & i \xi^{g1}_{HCM} & 0 \\
\vspace{-12pt}
&& \\
-i \xi^{g2}_{HCM} & \xi^y_{HCM} & 0 \\
\vspace{-12pt}
&& \\
0 & 0 & \xi^z_{HCM}
\end{array}
\ri
\\ \vspace{-8pt} & \\ - i \sqrt{\epso
\muo} \,
  \le
\begin{array}{ccc} \xi^x_{HCM} &  i \xi^{g2}_{HCM} & 0 \\
\vspace{-12pt}
&& \\
-i \xi^{g1}_{HCM} & \xi^y_{HCM} & 0 \\
\vspace{-12pt}
&& \\
0 & 0 & \xi^z_{HCM}
\end{array}
\ri & \muo \,
  \le
\begin{array}{ccc} \mu^x_{HCM} & i \mu^g_{HCM} & 0 \\
\vspace{-12pt}
&& \\
-i \mu^g_{HCM} & \mu^y_{HCM} & 0 \\
\vspace{-12pt}
&& \\
0 & 0 & \mu^z_{HCM}
\end{array}
\ri
\end{array} \ris, \l{K_HCM_fcm2}
\end{equation}
which is rather more general than the form \r{K_HCM_fcm1}.

\begin{figure}[!ht]
\centering \psfull
   \epsfig{file=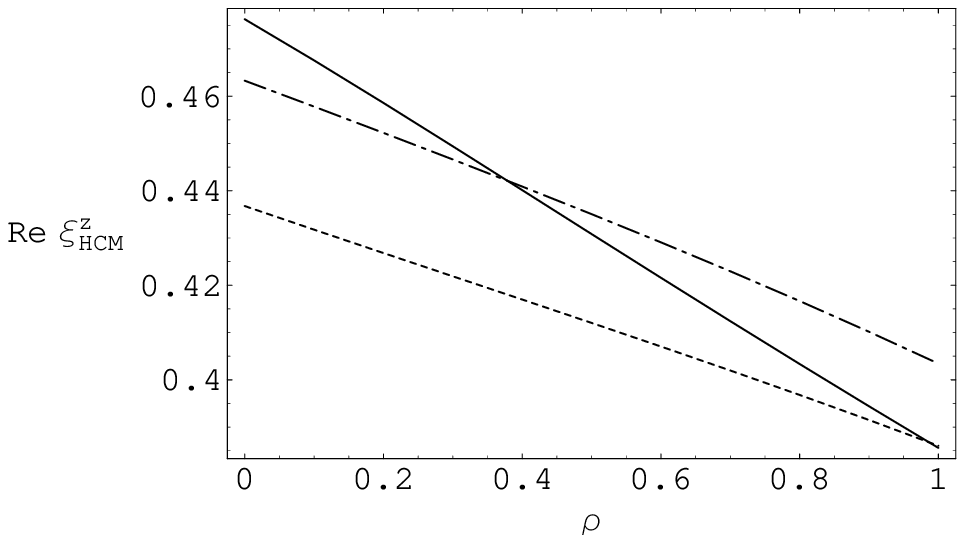,width=3.0in} \hfill
  \epsfig{file=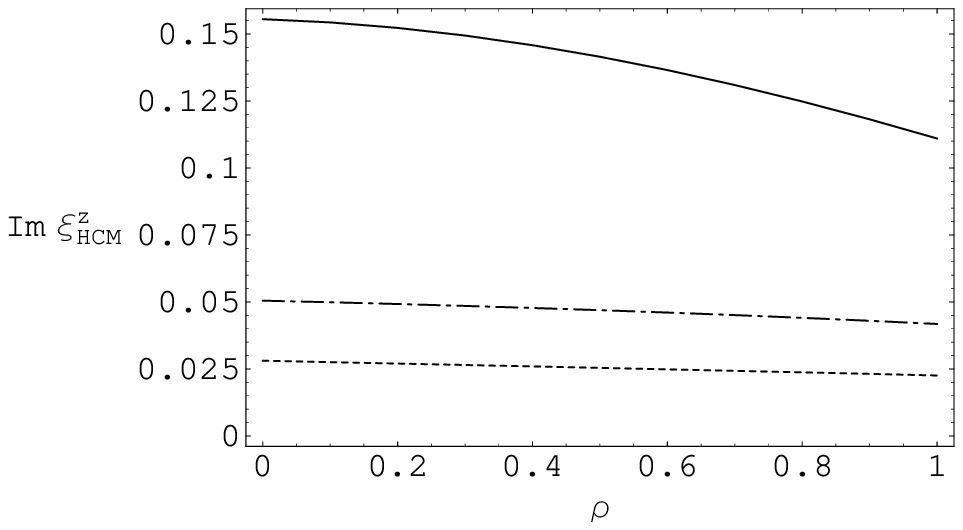,width=3.0in}
\caption{ \label{fig6} Real (left) and imaginary (right) parts of
the HCM constitutive parameter $\xi^{z}_{HCM}$  plotted against the
eccentricity parameter $\rho \in (0,1)$  for $ \eta / \lo = 0$
(dashed curves), $ \eta / \lo =   0.05$ (broken dashed curves)
 and $ \eta / \lo =  0.1 $ (solid curves). The HCM is a Faraday
 chiral medium.
}
\end{figure}

As a representative example, graphs of the real and imaginary parts
of
 $\xi^{z}_{HCM}$  versus
 $\rho \in (0,1) $ are exhibited in Figure~\ref{fig6} for $ \eta / \lo \in \lec 0, 0.05, 0.1 \ric$.
The real part of $\xi^{z}_{HCM}$ is a strong function of $\rho$,
whereas the imaginary part varies  less as $\rho$ increases. The
sensitivity of both the real and imaginary parts of $\xi^{z}_{HCM}$
to $\rho$ is clearly influenced by the value of $\eta$. Patterns of
behaviour with respect to $\rho$ that are qualitatively similar to
those displayed in Figure~\ref{fig6} are found for the HCM
constitutive parameters $\eps^{x,y,z,g}_{HCM}$,
$\xi^{x,y,g1,g2}_{HCM}$ and $\mu^{x,y,z,g}_{HCM}$ which are not
represented in Figure~\ref{fig6}.

\subsubsection{Particle size and particle
 orientation}

\begin{figure}[!ht]
\centering \psfull \epsfig{file=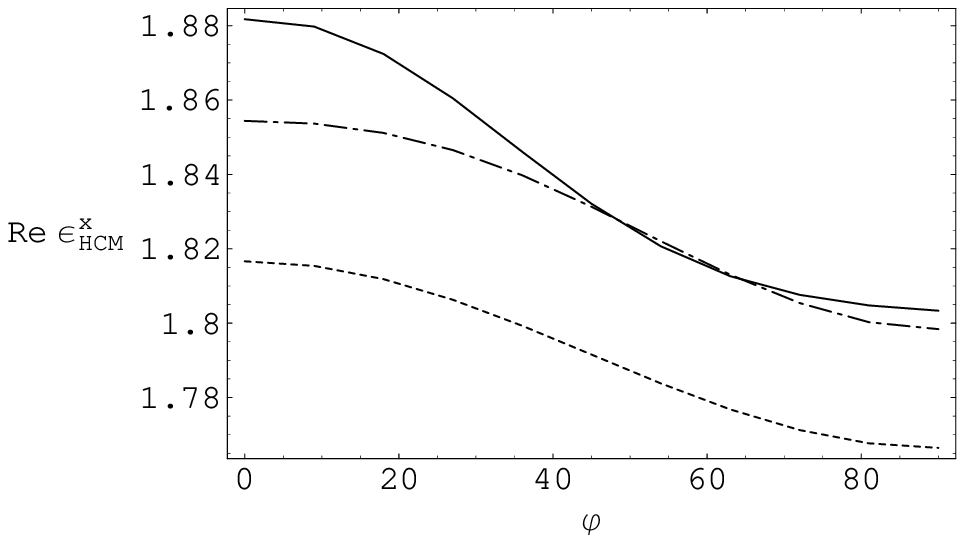,width=3.0in} \hfill
  \epsfig{file=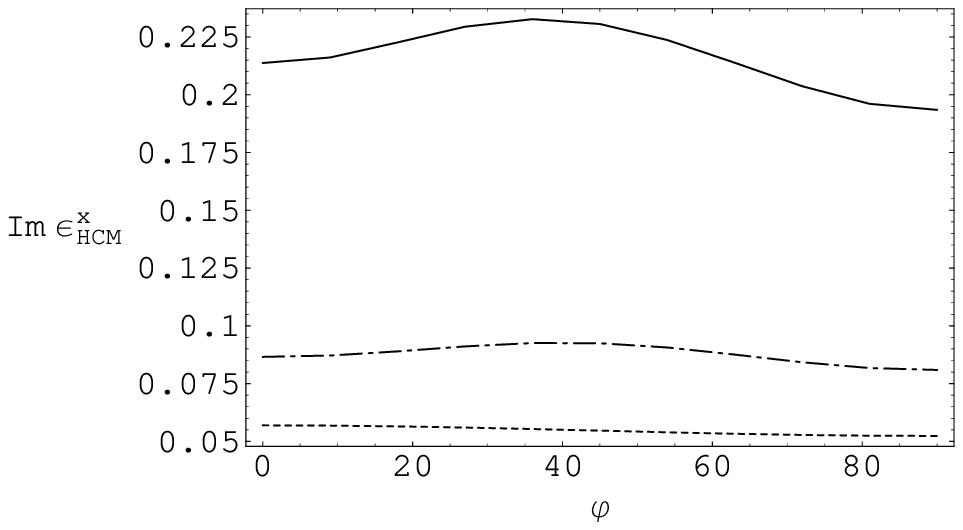,width=3.0in}  \\
   \epsfig{file=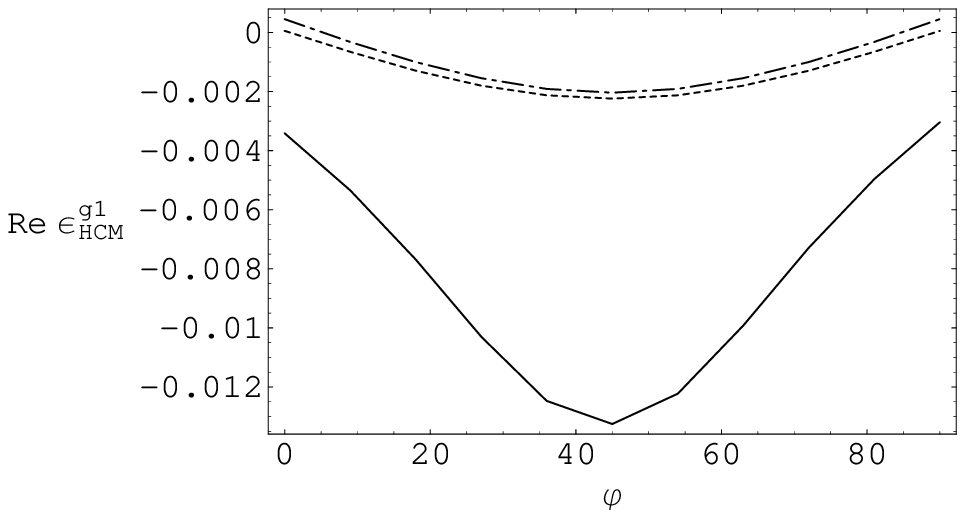,width=3.0in} \hfill
  \epsfig{file=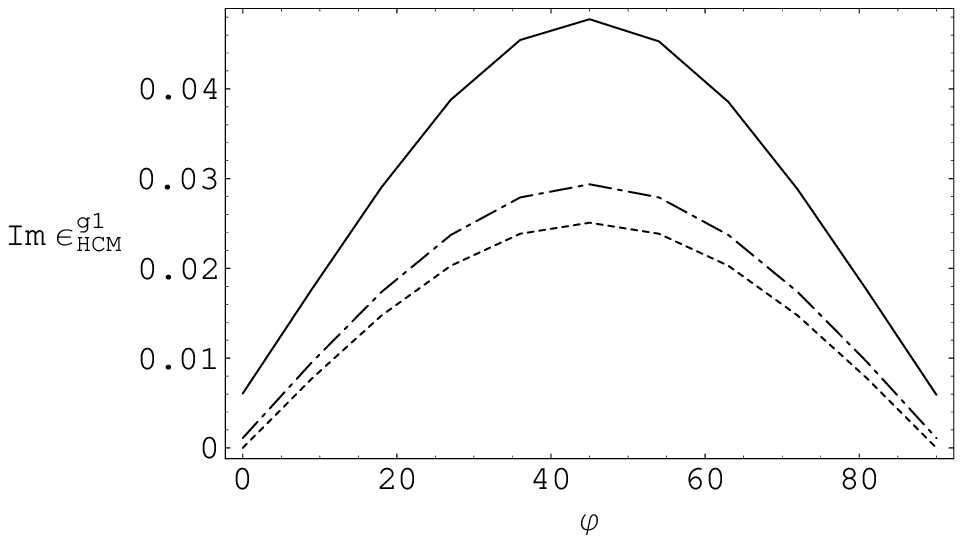,width=3.0in}
\caption{ \label{fig7} Real (left) and imaginary (right) parts of
the HCM constitutive parameters $\eps^{x,g1}_{HCM}$
  plotted against
 orientation angle $\varphi \in (0,\pi/2)$ for $ \eta / \lo = 0$ (dashed curves), $ \eta / \lo =   0.05$ (broken dashed curves)
 and $ \eta / \lo =  0.1 $ (solid curves). The HCM is a Faraday
 chiral medium.
}
\end{figure}

Now we turn to  effect of particle size  $\eta$ in relation to
particle orientation. Let us fix the following parameters:  volume
fraction $f_a = 0.5$, particle eccentricity $\rho = 1$ and the
correlation length $L = 0$. Consequently,  the HCM constitutive
dyadic has the form
\begin{equation}
\*K_{\,HCM} = \les \begin{array}{cc} \epso \,  \le
\begin{array}{ccc} \eps^x_{HCM} & i \eps^{g1}_{HCM} & 0 \\
\vspace{-12pt}
&& \\
-i \eps^{g2}_{HCM} & \eps^y_{HCM} & 0 \\
\vspace{-12pt}
&& \\
0 & 0 & \eps^z_{HCM}
\end{array}
\ri
 & i \sqrt{\epso
\muo} \,
  \le
\begin{array}{ccc} \xi^x_{HCM} & i \xi^{g1}_{HCM} & 0 \\
\vspace{-12pt}
&& \\
-i \xi^{g2}_{HCM} & \xi^y_{HCM} & 0 \\
\vspace{-12pt}
&& \\
0 & 0 & \xi^z_{HCM}
\end{array}
\ri
\\ \vspace{-8pt} & \\ - i \sqrt{\epso
\muo} \,
  \le
\begin{array}{ccc} \zeta^x_{HCM} &  i \zeta^{g1}_{HCM} & 0 \\
\vspace{-12pt}
&& \\
-i \zeta^{g2}_{HCM} & \zeta^y_{HCM} & 0 \\
\vspace{-12pt}
&& \\
0 & 0 & \xi^z_{HCM}
\end{array}
\ri & \muo \,
  \le
\begin{array}{ccc} \mu^x_{HCM} & i \mu^{g1}_{HCM} & 0 \\
\vspace{-12pt}
&& \\
-i \mu^{g2}_{HCM} & \mu^y_{HCM} & 0 \\
\vspace{-12pt}
&& \\
0 & 0 & \mu^z_{HCM}
\end{array}
\ri
\end{array} \ris, \l{K_HCM_fcm3}
\end{equation}
which is more general than \r{K_HCM_fcm1} and \r{K_HCM_fcm2}. As
illustrative   examples, the
  the real and imaginary parts of
$\eps^{x,g1}_{HCM}$ are plotted in Figure~\ref{fig7} against
 $\varphi \in (0,\pi/2) $ for $ \eta / \lo \in \lec 0, 0.05, 0.1 \ric$.
The particle size $\eta$ has a strong influence on the real and
imaginary parts of $\eps^{g1}_{HCM}$, as well as on the imaginary
part of $\eps^{x}_{HCM}$. The influence on the real part of
$\eps^{x}_{HCM}$ is smaller by comparison, but still significant.
The graphs for $\eps^{g1}_{HCM}$ are symmetric about the line
$\varphi = \pi/4$ whereas those for $\eps^{x}_{HCM}$ are not.
Broadly similar behaviour is exhibited by the HCM constitutive
parameters not plotted in Figure~\ref{fig7}, namely,
$\eps^{y,z,g2}_{HCM}$, $\xi^{x,y,z,g1,g2}_{HCM}$,
$\zeta^{x,y,g1,g2}_{HCM}$ and $\mu^{x,y,z,g1,g2}_{HCM}$.

\subsubsection{Particle size and correlation length}

Finally, to focus upon the effect of particle size  $\eta$ in
relation to correlation length $L$,  the volume fraction is fixed at
$f_a = 0.5$, the orientation angle at  $\varphi = 0$ and the
eccentricity parameter at $\rho = 0$. The corresponding HCM
constitutive dyadic has the form \r{K_HCM_fcm1}.

\begin{figure}[!ht]
\centering \psfull \epsfig{file=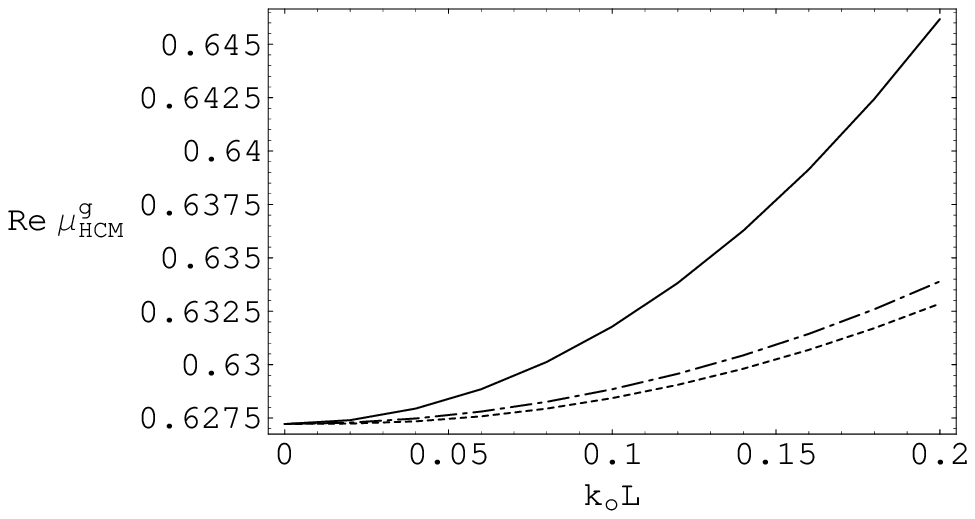,width=3.0in} \hfill
  \epsfig{file=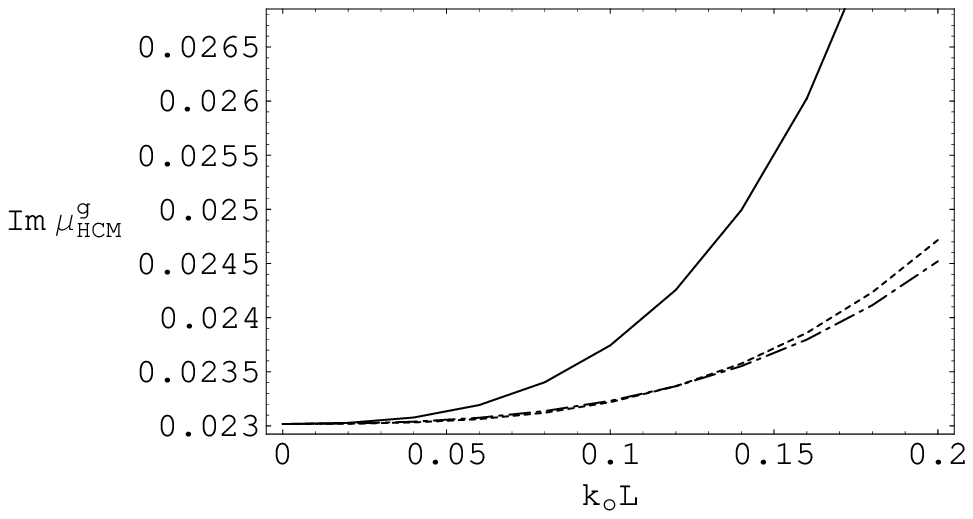,width=3.0in}
\caption{ \label{fig8} Real (left) and imaginary (right) parts of
the HCM constitutive parameter $\mu^{g}_{HCM}$  plotted against
 relative correlation length
  $\ko L  \in (0,0.2)$ for $ \eta / L = 0$ (dashed curves), $ \eta / L =   0.5$ (broken dashed curves)
 and $ \eta / L =  0.95 $ (solid curves). The HCM is a Faraday
 chiral medium.
}
\end{figure}

In Figure~\ref{fig8}, the real and imaginary parts of
$\mu^{g}_{HCM}$ are plotted against $\ko L  \in (0,0.2)$ for $ \eta
/ L \in \lec 0, 0.5, 0.95 \ric$. The pattern of behaviour with
respect to $L$ is similar to that presented in Figure~\ref{fig4} for
the biaxial binisotropic HCM. That is, the imaginary part of
$\mu^{g}_{HCM}$ is more obviously sensitive to inceasing $L$ than is
the real part. In addition, both the real and imaginary parts of
$\mu^{g}_{HCM}$ are more sensitive to $L$ at larger values of
$\eta$. The other HCM constitutive parameters, namely
$\eps^{x,z,g}_{HCM}$, $\xi^{x,z,g}_{HCM}$ and $\mu^{x,z}_{HCM}$,
respond in a generally similar manner as $L$ increases for the three
values of $\eta$ considered here.

\section{Concluding remarks} \l{conc_remarks}

Homogenization formalisms, such as  the widely-used Maxwell Garnett
and Bruggeman formalisms,  often inadequately take into account the
distributional statistics and sizes of the component phase
particles. The SPFT~---~through describing the distributional
statistics of the component phases in terms of a hierarchy of
spatial correlation functions~---~provides  a
  conspicuous
exception. In the preceding sections,  an extension to the SPFT for
the most general linear class of HCM is developed, in which
 a nonzero volume  is attributed
to the component phase particles. By means of extensive numerical
calculations, based on Lorentz--reciprocal and
Lorentz--nonreciprocal HCMs, it is demonstrated that estimates of
the HCM constitutive parameters in relation to volume fraction,
particle eccentricity, particle orientation and correlation length
are all significantly influenced by the size of the component phase
particles.

It is particularly noteworthy that the influence of the particle
size is generally stronger on the imaginary parts of the HCM
constitutive parameters than it is on the corresponding real parts.
 In this respect, the effect of $\eta$ is reminiscent of the
effect of the correlation length in the second order SPFT
\c{MLW_AEU}. Increasing the correlation length for the second order
SPFT generally results in an increase in the degree of dissipation
associated with the HCM. This dissipative loss is attributed to
radiative scattering losses from the macroscopic coherent field
\c{MLW00,Kranendonk}.
 It may be observed in Figures~4 and 8 (and in numerical results for
 other HCM constitutive parameters
 not presented here) that
the effects of particle size and correlation length on the estimates
of the imaginary parts of the HCM constitutive parameters are
generally cumulative. This suggests that  coherent scattering losses
associated with the HCM become greater as
 the  particle size increases. A similar finding was reported for
 anisotropic dielectric HCMs \c{M04}.

 In conclusion, the importance of incorporating microstructural
details, such component particle size and spatial distribution,
within homogenization formalisms is further emphasized by this
study.

\vspace{10mm} \noindent {\bf Acknowledgement:} JC is supported by a
Scottish Power--EPSRC  Dorothy  Hodgkin Postgraduate Award.

\newpage

\newpage

\newpage

\newpage

\newpage

\newpage

\newpage

\newpage

\end{document}